\documentclass[10pt,conference]{IEEEtran}
\usepackage[margin=0.7in]{geometry}
\usepackage{times}
\usepackage{amsmath,amssymb}
\usepackage{graphicx}
\usepackage{caption}
\usepackage{subcaption}
\usepackage{siunitx}
\usepackage{authblk}
\usepackage{url, balance}  

\usepackage[dvipsnames]{xcolor}
\usepackage{orcidlink}

\usepackage{fancyhdr}
\fancypagestyle{firstpage}{
    \fancyhf{}
    \fancyhead[C]{\footnotesize
    This work has been submitted to the IEEE for possible publication.
    Copyright may be transferred without notice, after which this version
    may no longer be accessible.}
    
}

\title{Doppler Effect in High-Mobility Free Space Optical Links with Multicarrier Intensity Modulation}

\hypersetup{
    colorlinks=false,
    hidelinks=true,
}

\author[1]{M. H. Abid\orcidlink{0000-0002-7799-9807}}
\author[1]{M. A. Khalighi\orcidlink{0000-0002-0522-6688}} 
\author[2]{M. Safari\orcidlink{0000-0001-7777-0052}} 
\affil[1]{Aix-Marseille University, CNRS, Centrale Med, Fresnel Institute, Marseille, France} 
\affil[2]{School of Engineering, the University of Edinburgh, Edinburgh EH9 3JL,
UK} 
\date{}

\begin{document}
\maketitle
\thispagestyle{firstpage}

\begin{abstract}
In this paper, we investigate the impact of Doppler-induced time scaling in intensity-modulation/direct-detection (IM/DD) multicarrier free-space optical links subject to high mobility. Modeling the Doppler effect as a single-sideband frequency shift applied to the analytic signal waveform, we show that it produces a baseband-equivalent phase rotation, resulting in a common-phase error (CPE), in addition to Dirichlet-kernel inter-carrier interference (ICI), even though the optical carrier phase is removed through photo-detection. The numerical results based on DC-biased optical orthogonal frequency-division multiplexing (DCO-OFDM), show that at small Doppler shifts, the CPE component dominates and can be effectively compensated through pilot-aided phase tracking, while the induced ICI remains rather negligible. At larger Doppler shifts, however, ICI becomes dominant and results in a bit-error-rate floor, highlighting the need for both CPE tracking and ICI mitigation. This is particularly relevant for optical links with small subcarrier spacing and large Doppler shifts, such as those in low-Earth-orbit satellite-to-ground and inter-satellite transmission scenarios.

\end{abstract}

\section{Introduction}
\label{sec:intro}
Free-space optical (FSO) communication is gaining significant attention as a key enabler to meet next-generation (NextG) wireless connectivity requirements~\cite{toyoshima-JLT-2021}. Coherent modulation and/or detection provide improved performance, notably through increased receiver (Rx) sensitivity, but at the expense of higher system complexity and cost~\cite{Fernandes-ComMag-2023, guiomar-JLT-2022}. By contrast, noncoherent signaling based on intensity-modulation and direct-detection (IM/DD) remains the conventional approach due to its implementation simplicity~\cite{Khalighi-SomSurvey-2014}. 

\begin{figure*}[t]
  \centering

  \begin{subfigure}{\linewidth}
    \centering
    \includegraphics[scale=.477]{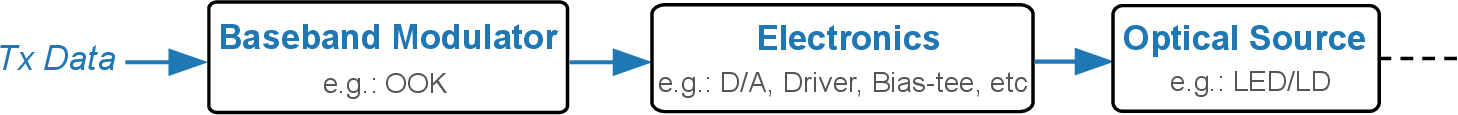}
    \caption{Single-subcarrier Tx}
    \label{fig:stacked-a}
  \end{subfigure}

  \vspace{3mm} 

  \begin{subfigure}{\linewidth}
    \centering
    \includegraphics[scale=.477]{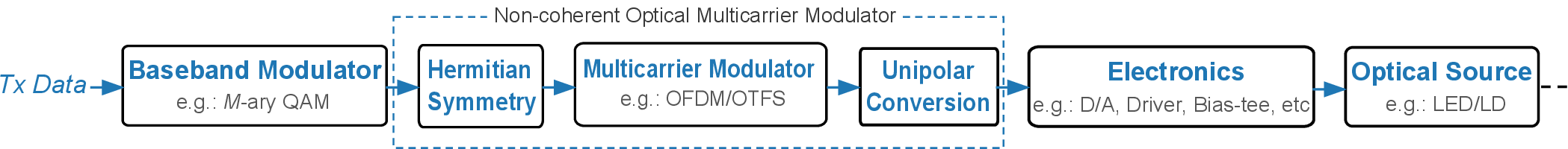}
    \caption{Multiple-subcarrier Tx}
    \label{fig:stacked-b}
  \end{subfigure}

  \caption{{Tx block diagrams for IM/DD based systems. (a) Single-subcarrier Tx; (b) multiple-subcarrier Tx, where QAM symbols are mapped with Hermitian symmetry, followed by unipolar conversion (e.g., DC bias adding and zero-level clipping).} 
\label{fig:tx}
}
  \label{fig:stacked}
\end{figure*}

In the IM/DD systems, the photo-detector (PD) converts the received optical intensity into photo-current through square-law detection, thereby discarding the optical carrier phase. Conventional systems rely on baseline constellations such as on–off keying (OOK), resulting in an effectively single-subcarrier modulation scheme, as shown in Fig.\,\ref{fig:stacked-a}. 

To use larger signal constellations such as quadrature amplitude modulation (QAM), multiple-subcarrier schemes such as 
optical orthogonal frequency-division multiplexing (O-OFDM) or optical orthogonal time–frequency space (O-OTFS) are often employed. In this case, Hermitian symmetry is imposed on the complex-valued constellation symbols so that the subsequent inverse discrete Fourier transform (IDFT) produces a real-valued time-domain waveform, as shown in Fig.~\ref{fig:stacked-b}~\cite{Armstrong-JLT-2013, Chen-PTL-2025}.
At the Rx, after the discrete Fourier transform (DFT), the complex QAM symbols are reconstructed.  
Here, it is important to note that, although the optical-carrier phase is lost through photo-detection, the baseband signal phase is preserved due to the complex nature of the symbols~\cite{Morelli-IEEE-Proceedings-2007}. Consequently, coherent-like channel effects reappear in the electrical (baseband) domain~\cite{Schmidl-TCOM-2002}.
In other words, any channel effect that perturbs the signal in the frequency domain (such as Doppler shift) manifests through the electrical subcarrier frequencies and distorts the baseband phase. 

As a related finding, experimental studies in~\cite{Maho-ICSO-2019} reported negligible Doppler impact in an OOK-based inter-atmospheric IM/DD link, whereas a clear Doppler effect appeared for differential phase-shift keying (DPSK) with a delay-line interferometer (DLI). The reason is that the DLI mixes optical fields with different phases before photo-detection, resulting in phase-dependent interference. In line with this, in satellite IM/DD links such as low-Earth-orbit (LEO) to optical-ground-station (OGS) and inter-satellite links (ISL), Doppler-induced frequency offsets are conventionally neglected since square-law detection removes the optical-carrier phase~\cite{wang-COM_Mag-2024, giggenbach-wiley-2023}.

Motivated by the above discussion, we distinguish the optical-carrier phase from the \emph{baseband-equivalent (BE) phase}, whose evolution is affected by frequency-dependent impairments such as timing offsets and Doppler shift. A key contribution of this paper is the formalization of this BE phase and the demonstration that, in multicarrier IM/DD links, Doppler-induced time scaling leads (after direct detection and DFT-based demodulation) to a nontrivial impairment structure in the electrical domain. In particular, we show how this effect manifests as a common phase rotation combined with Dirichlet-kernel inter-carrier interference (ICI), despite the removal of the optical-carrier phase by the photo-detector. These impairments motivate pilot-aided common-phase error (CPE) tracking and, when necessary, ICI-aware equalization, unlike in conventional single-subcarrier IM/DD links~\cite{Maho-ICSO-2019, wang-COM_Mag-2024, giggenbach-wiley-2023}. To the best of our knowledge, this Doppler-induced CPE/ICI manifestation has not been explicitly characterized before for multicarrier IM/DD systems in high-mobility scenarios.

Note that the use of FSO links in high-mobility scenarios such as LEO-to-OGS and ISL is motivated by several advantages, including ultra-high capacity, superior energy efficiency, enhanced security, and immunity to RF spectrum congestion~\cite{toyoshima-JLT-2021, Khalighi-SomSurvey-2014}.
Here, multicarrier signaling is further motivated by several benefits, including more efficient turbulence mitigation~\cite{Djordjevic-OE-2008}, the ability to overcome transmitter (Tx)/receiver (Rx) frontend bandwidth limitations, and simplified timing recovery and equalization in high-data-rate systems~\cite{mrabet_WileyTransaction_2022}.

The rest of the paper is organized as follows.  Section~\ref{sec:model} provides a mathematical analysis of the Doppler effect in multicarrier IM/DD systems. This is validated via numerical results in Section~\ref{sec:results}. Lastly, Section~\ref{sec:disc} summarizes the key insights, discusses the implications, and concludes the paper.

%

\section{System Model} 
\label{sec:model} 
We consider an IM/DD line-of-sight (LOS) FSO link with time-varying propagation delay due to the relative motion between the Tx and Rx platforms, as it can be the case in the link between a LEO satellite and an OGS or in LEO ISLs. Atmospheric turbulence effects are neglected to isolate the Doppler effect. 
We consider DC-biased optical (DCO)-OFDM signaling based on $N$-point IDFT and Hermitian symmetry applied to QAM symbols, i.e., $X[N-k]=X^{\ast}[k]$ and $X[0]=X[N/2]=0$, $k \in \{1,\dots,N/2-1\}$. The resulting time-domain signal is then~\cite{Armstrong-JLT-2013}:
\begin{equation} \label{eq:mod}
x[n]=\frac{1}{N}\sum_{k=0}^{N-1} X[k]\; e^{j2\pi kn/N}, \hspace{0.5em} n=0,\dots,N-1.
\end{equation}
To ensure non-negativity of the resulting signal, we add a bias $b$ to $x[n]$ and zero-clip the signal, while neglecting clipping effects, including those related to the optical source’s limited dynamic range~\cite{Dimitrov-Haas-TCOM-2012}.
The resulting real, non-negative electrical waveform is then used to drive the optical source. We further define the zero-mean (DC-bias-free) O-OFDM signal by $x_s(t)$, i.e., $x_s(t)=x(t)-b$. For simplicity, and in order to focus on the Doppler effect, we neglect the channel loss and pointing-error effects, and consider normalized channel coefficient. Then, based on the received signal $y(t)$, we define $y_s(t)=y(t)-b$, which accounts for both the effects of Doppler shift and the Rx noise $w(t)$, modeled as additive white Gaussian noise of variance $\sigma_w^2$. 

\subsection{Doppler shift modeling}\label{Subsec-Doppler-Mod}
 
Let $R(t)$ denote the Tx–Rx range, resulting in the propagation delay $\tau(t)=R(t)/c$. For constant radial velocity $v$~\cite{Hao-Radioengineering-2023},
\begin{equation} \label{eq:D}
    R(t)=R_0 \mp v t,
\end{equation}
{where the negative sign corresponds to an approaching Tx–Rx pair (decreasing range),
and the positive sign to a receding pair (increasing range). Here $R_0$ is the initial range at
$t=0$.} For an approaching Tx–Rx pair, the corresponding delay is:
\[
\tau(t)=\tau_0 - \frac{v}{c}t \ \ ,\qquad \tau_0 \triangleq \frac{R_0}{c}.
\]
In other words,
\begin{equation*}
\begin{aligned}
y(t) &= h\, x\big(t-\tau(t)\big) = b + x_s\big((1+\alpha)t-\tau_0\big), 
\end{aligned}
\end{equation*}
where $\alpha\triangleq v/c$. 
Thus, the zero-mean waveform $x_s(t)$ is time-scaled by the
factor $(1+\alpha)$. In the frequency domain, this is equivalent 
to multiplying all non-zero subcarrier frequencies by  $(1+\alpha)$. That is, each
subcarrier at frequency $f_i$ is shifted to $(1+\alpha)f_i$,
resulting in a frequency offset of $\alpha f_i$. 
Instead of considering the exact Doppler shift for each subcarrier, here we adopt the approximation of considering a fixed Doppler shift $f_D$ evaluated at a nominal frequency $f_m$, i.e., $f_D \triangleq \alpha f_m$~\cite{boashash_AcademicPress_2015}. This is a commonly-used approach; for instance, for multi-carrier RF systems, Doppler is typically evaluated at the central carrier frequency~\cite{matz-sig.proc.mag-2013, hong-Academic_Pres-2022}. Here, $f_m$ can be chosen as the center subcarrier frequency (representing hence the average Doppler across subcarriers) or as the band-edge subcarrier frequency (equivalent of considering a worst-case Doppler assumption for all subcarriers).\footnote{For an $N$-subcarrier DCO-OFDM system with subcarrier spacing $\Delta f$, the nominal frequency may be chosen as $f_{m,\mathrm{c}} =  (N/4)\Delta f$ for a band-center reference, or $f_{m,\mathrm{e}} = (N/2-1)\Delta f$ for the band-edge subcarrier frequency, where the indices refer to the positive-frequency data-bearing subcarriers.} 
In this study, we adopt the second approach to define a single worst-case reference Doppler shift, which represents a standard conservative assumption when quantifying Doppler severity in multicarrier ICI analysis~\cite{Kaiser-VTC-1999, montalban-IET-2014}.

To emulate this equivalent Doppler-induced frequency shift $f_D$ over the duration of one OFDM symbol, we apply
a single-sideband (SSB) frequency shift only to the zero-mean part of the received signal, i.e., $y_{s}[n]$ \cite{Zaman-SIBIRCON-2010, vanDeBeek_TSP_1997}. This is done using the analytic
signal of $x[n]$ in (\ref{eq:mod}), denoted by $\tilde{x}[n]$, where
\begin{align}
y_{s}[n] &= \Re\!\left\{\tilde{x}[n]\; e^{j2\pi \varepsilon n}\right\}. \label{eq:ssb}
\end{align}
Here, $\Re\{\cdot\}$ denotes the real-part operator, and we define
\begin{align}
\varepsilon &\triangleq \frac{f_D}{F_s}=\frac{f_D}{N\Delta f}=\frac{\mu}{N},\qquad \text{with}\ \ 
\mu\triangleq \frac{f_D}{\Delta f}, \label{eq:epsdef}
\end{align}
where $F_s=N\Delta f$ is the sampling rate.
We note that in IM/DD links, we use  $f_D = \alpha f_m = \tfrac{v}{c}f_m$. 
In fact, Doppler-induced impairments in IM/DD multicarrier links are often negligible in low-mobility scenarios, but can become noticeable as mobility and bandwidth increase, as shown later in Section~\ref{sec:results}.


\begin{figure}
\captionsetup[subfigure]{justification=centering}
\centering

\begin{subfigure}[t]{0.48\columnwidth}
  \centering
  \includegraphics[width=\linewidth,height=3.5cm]{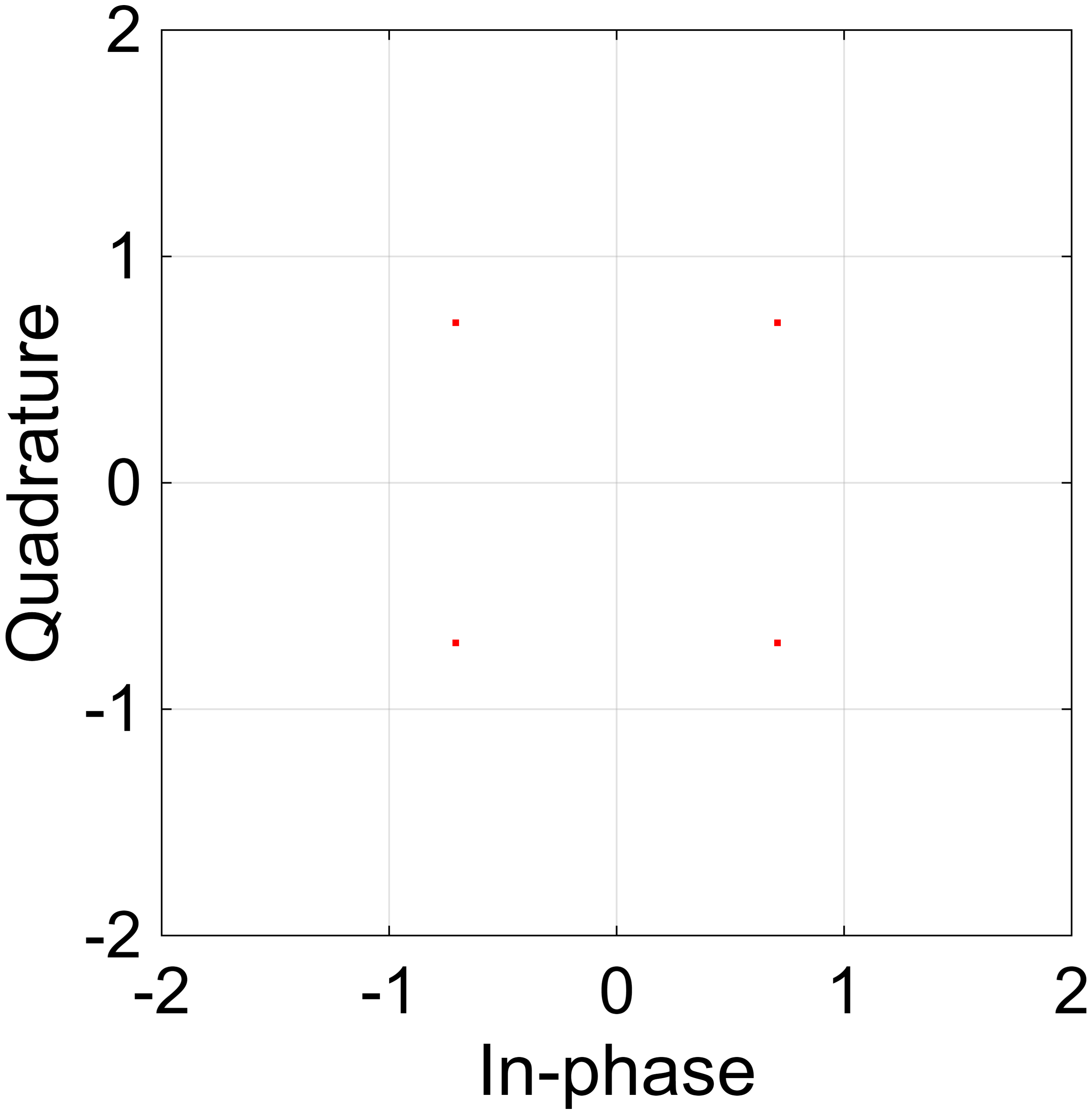}
  \caption{Transmitted symbols}
  \label{O2a}
\end{subfigure}\hfill
\begin{subfigure}[t]{0.48\columnwidth}
  \centering
  \includegraphics[width=\linewidth,height=3.5cm]{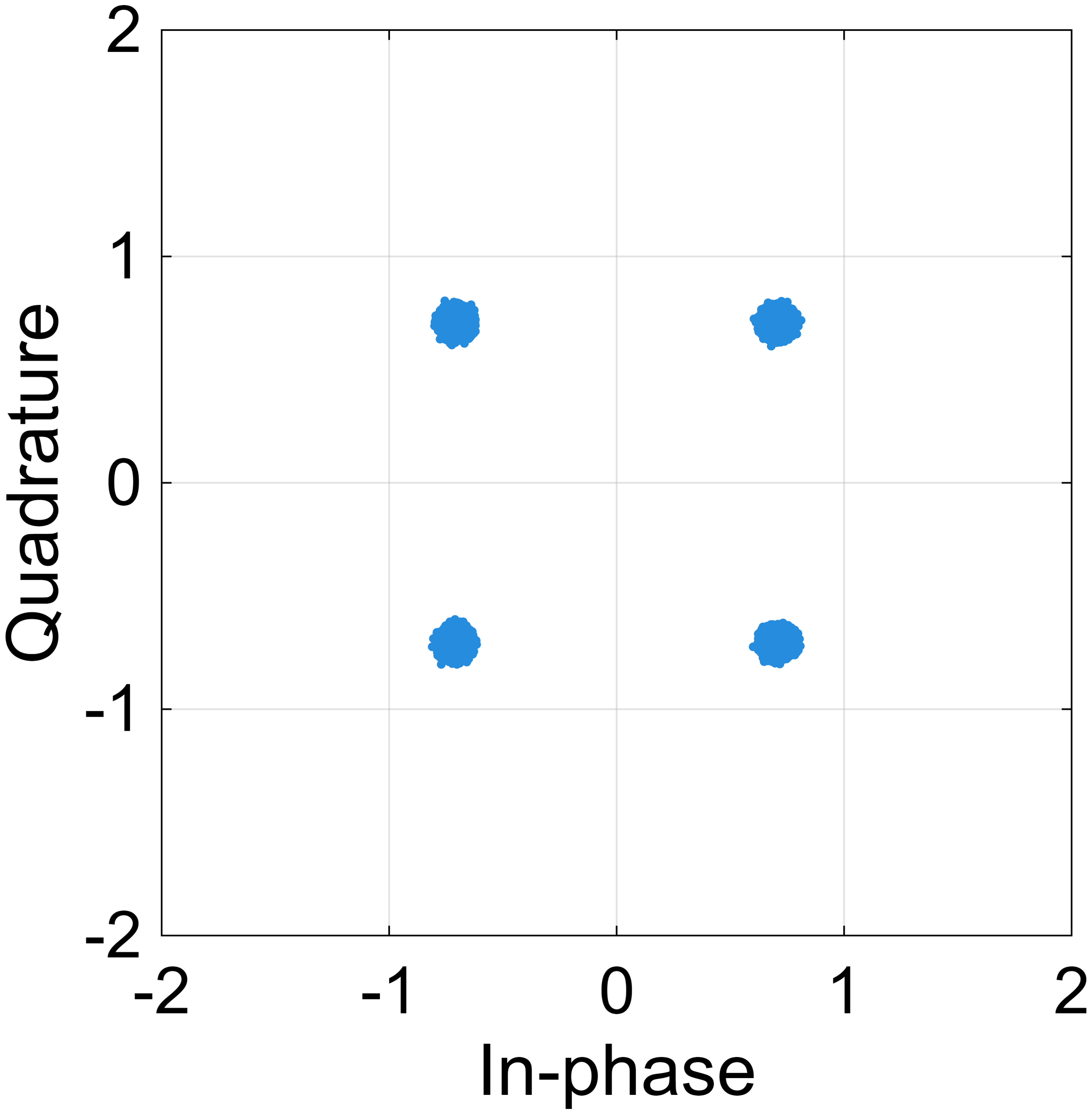}
  \caption{Received symbols, $\mu = 0$}
  \label{O2b}
\end{subfigure}

\medskip

\begin{subfigure}[t]{0.48\columnwidth}
  \centering
  \includegraphics[width=\linewidth,height=3.5cm]{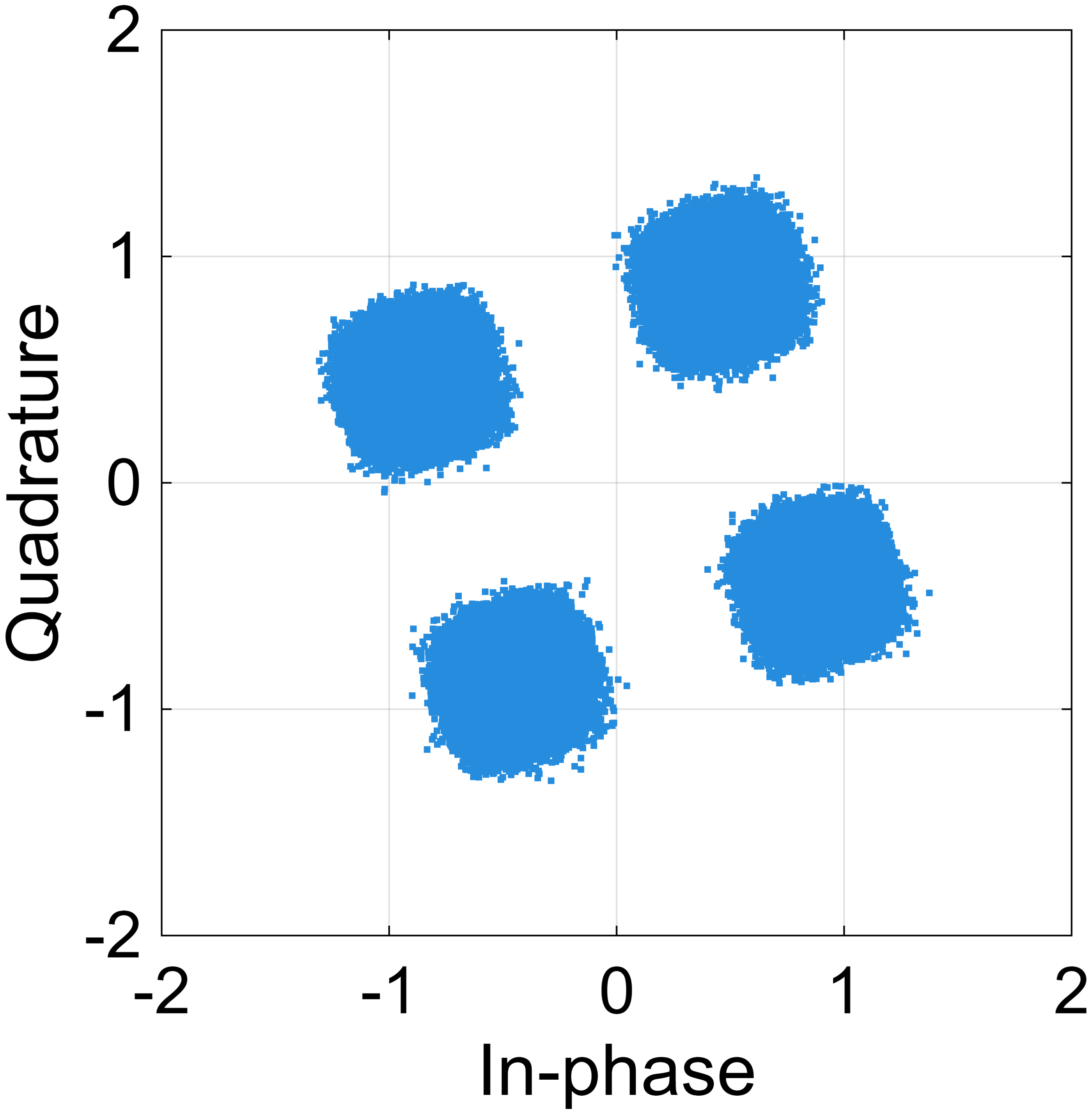}
  \caption{Received symbols, $\mu = 0.1$}
  \label{O2c}
\end{subfigure}\hfill
\begin{subfigure}[t]{0.48\columnwidth}
  \centering
  \includegraphics[width=\linewidth,height=3.5cm]{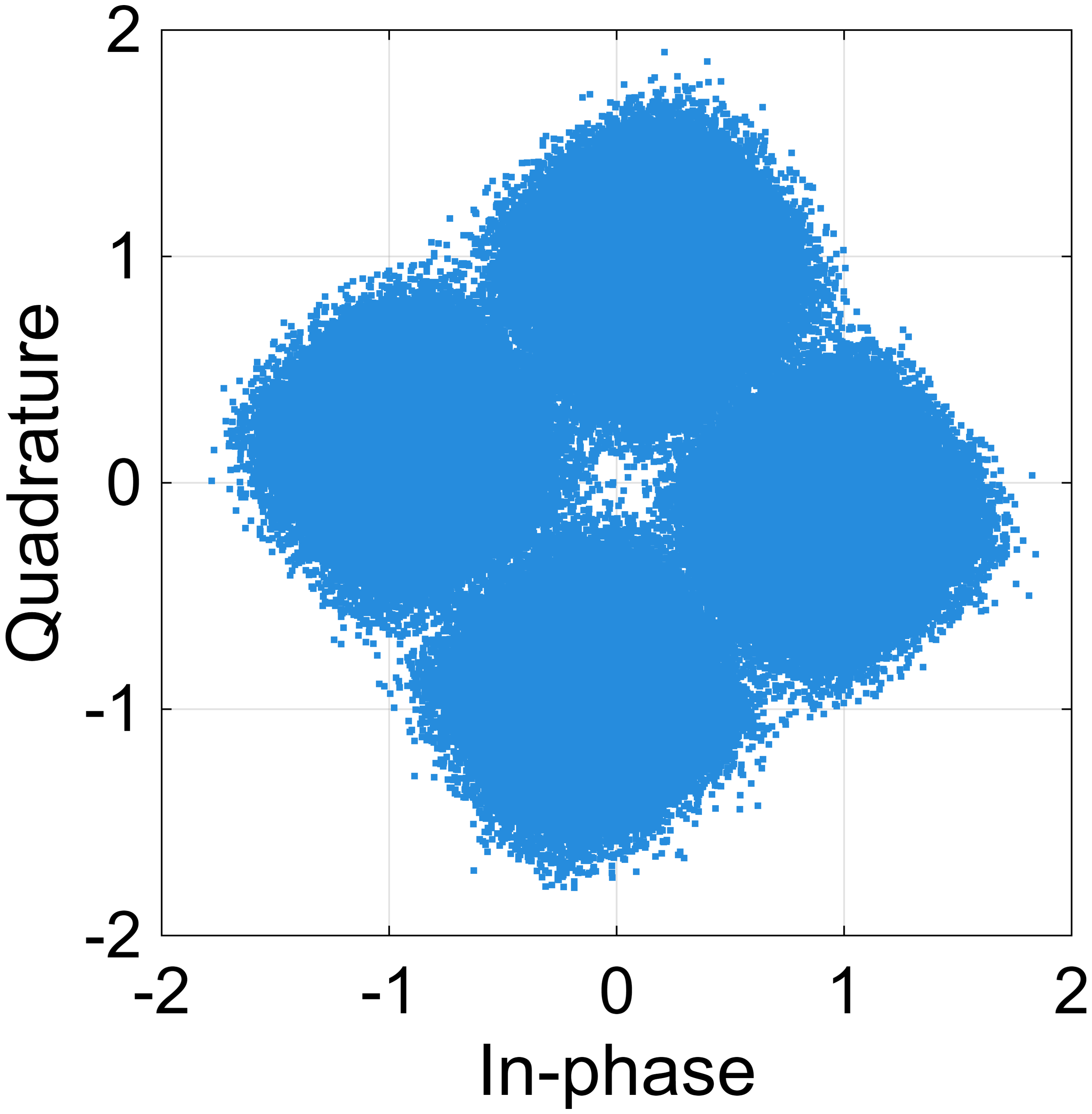}
  \caption{Received symbols, $\mu = 0.2$}
  \label{O2d}
\end{subfigure}

\medskip

\begin{subfigure}[t]{0.48\columnwidth}
  \centering
  \includegraphics[width=\linewidth,height=3.5cm]{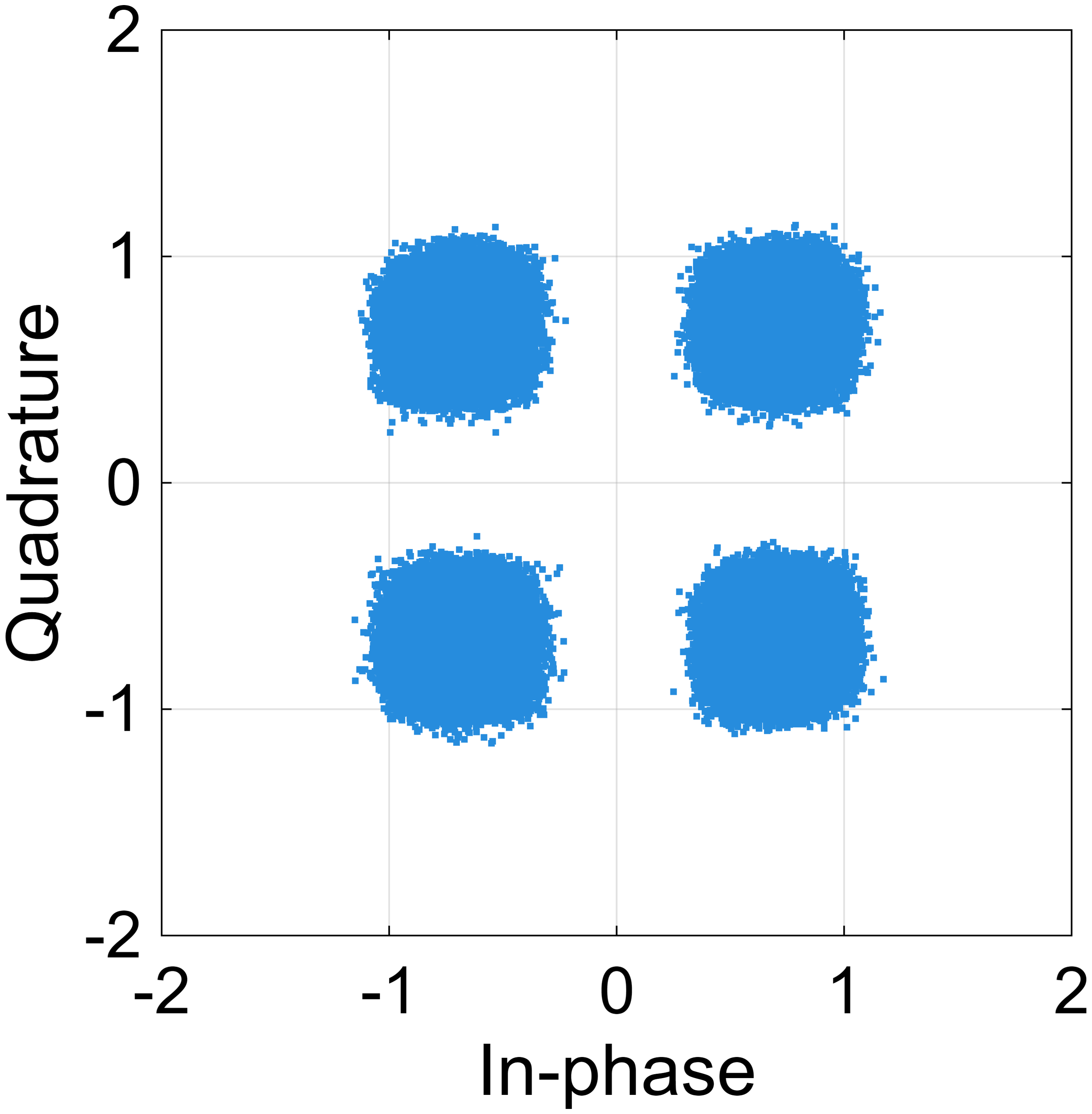}
  \caption{Received symbols with CPE correction, $\mu = 0.1$}
  \label{O2e}
\end{subfigure}\hfill
\begin{subfigure}[t]{0.48\columnwidth}
  \centering
  \includegraphics[width=\linewidth,height=3.5cm]{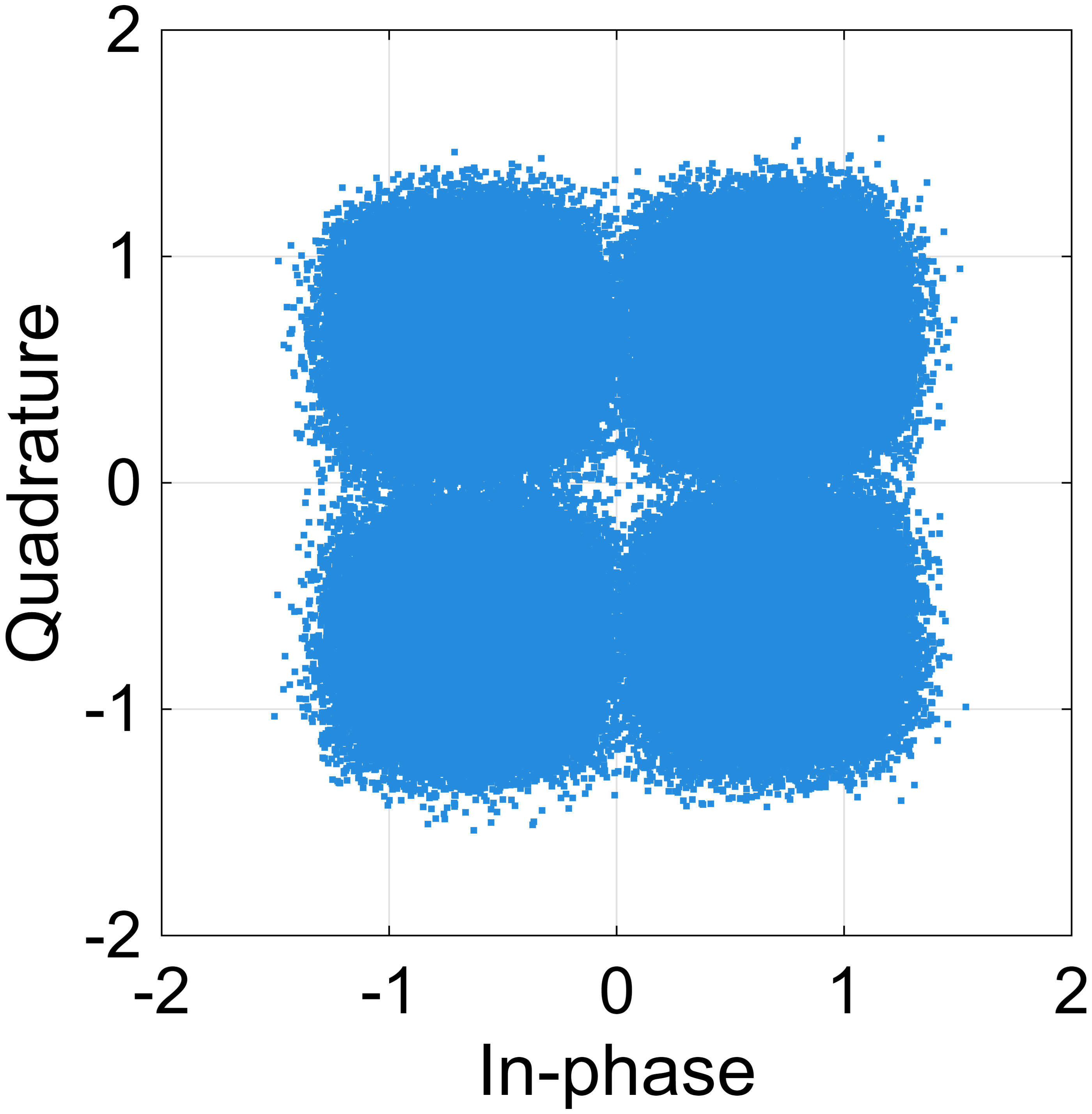}
  \caption{Received symbols with CPE correction, $\mu = 0.2$}
  \label{O2f}
\end{subfigure}

\caption{Signal constellations for 4-QAM-based DCO-OFDM; SNR = \SI{40}{dB}, {$N=4096$, $32$ pilot subcarriers.}}\label{R2}
\end{figure}

\begin{figure}[t]
    \centering

    \begin{subfigure}{\linewidth}
        \centering
        \includegraphics[width=\linewidth, height=0.15\textheight,keepaspectratio]{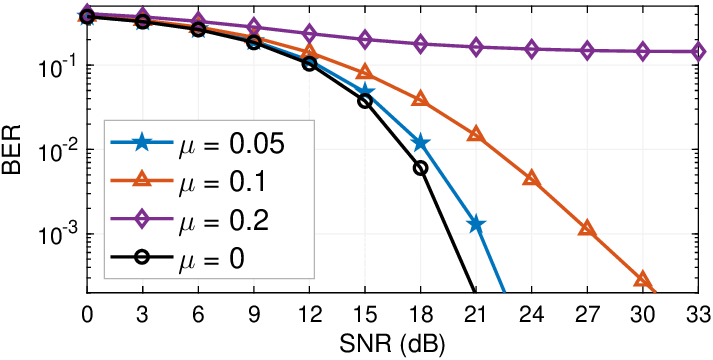}
        \caption{Before CPE correction}
        \label{BER1}
    \end{subfigure}

    \vspace{0.2cm}

    \begin{subfigure}{\linewidth}
        \centering
        \includegraphics[width=\linewidth, height=0.15\textheight,keepaspectratio]{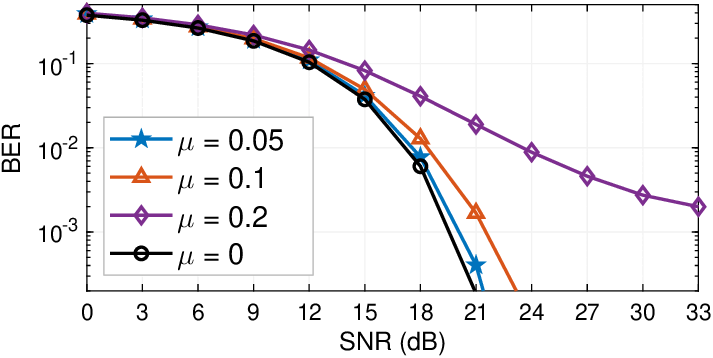}
        \caption{After CPE correction}
        \label{BER2}
    \end{subfigure}

    \caption{BER versus SNR for 4-QAM DCO-OFDM with different normalized Doppler $\mu$.}
    \label{BER}
\end{figure}


\subsection{Baseband Rx Model} \label{sec:rx-model}

Let us consider $y_s(t) = y_{s0}(t)+ w(t)$, where $y_{s0}(t)$ denotes noise-free DC-biased-removed received signal. We define the electrical SNR (excluding the DC bias) as
$\mathrm{SNR}=\mathbb{E}\{y_{s0}^2[n]\}/\sigma_w^2$. 
 Applying the $N$-point DFT to $y_s[n]$ (for OFDM demodulation) gives~\cite{Mokhtari-IET-2019}:
\begin{equation} 
Y_s[k] \;=\; \sum_{q=0}^{N-1} C_{k-q}(\varepsilon)\, X_s[q] \;+\; W[k], \quad k=0,\dots,N-1.
\label{eq:freqmodel}
\end{equation} 
where $W[k]$ is the DFT of the zero-mean AWGN Rx noise $w[n]$
and $C_{\ell}(\varepsilon)$ are the standard OFDM CFO coefficients (Dirichlet kernel). In the IM/DD framework, these latter appear as a baseband-equivalent CFO and are given by~\cite{Mokhtari-IET-2019}: 
\begin{equation}
C_{\ell}(\varepsilon)
=\frac{1}{N}\sum_{n=0}^{N-1} \exp\!\left\{j \frac{2\pi n}{N}\,\big(\varepsilon-\ell\big)\right\}, \qquad \ell\in\mathbb{Z}.
\end{equation}
When $\ell = k - q = 0$, the term $C_{0}(\varepsilon)$ contributes to a common phase rotation on each subcarrier due to the Doppler effect, known as CPE, while the $\ell \neq 0$ terms generate ICI, corresponding to contribution from other subcarriers $q\neq k$. 
On the other hand, for $\varepsilon = \mu= 0$ we have,
\begin{equation}
C_{\ell}(0)=\frac{1}{N} \sum_{n=0}^{N-1} e^{-j \frac{2\pi \ell}{N} n}
=
\begin{cases}
1, & \ell=0,\\
0, & \ell \neq 0,
\end{cases}
\end{equation}
i.e., $C_{\ell}(0) = \delta[\ell]$, where (\ref{eq:freqmodel}) reduces to $Y_s[k]=X_s[k]\ + W[k]$, which means the absence of CPE and ICI.

\begin{figure*}[!t]
\centering
\begin{subfigure}[t]{0.45\textwidth}
  \centering
  \includegraphics[width=\linewidth]{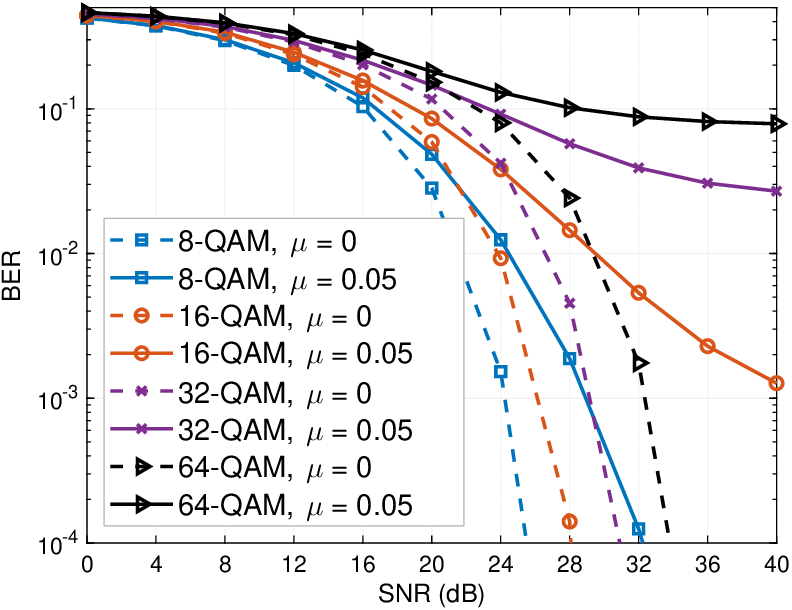}
  \caption{Before CPE correction.}
  \label{QAM1}
\end{subfigure}\hfill
\begin{subfigure}[t]{0.45\textwidth}
  \centering
  \includegraphics[width=\linewidth]{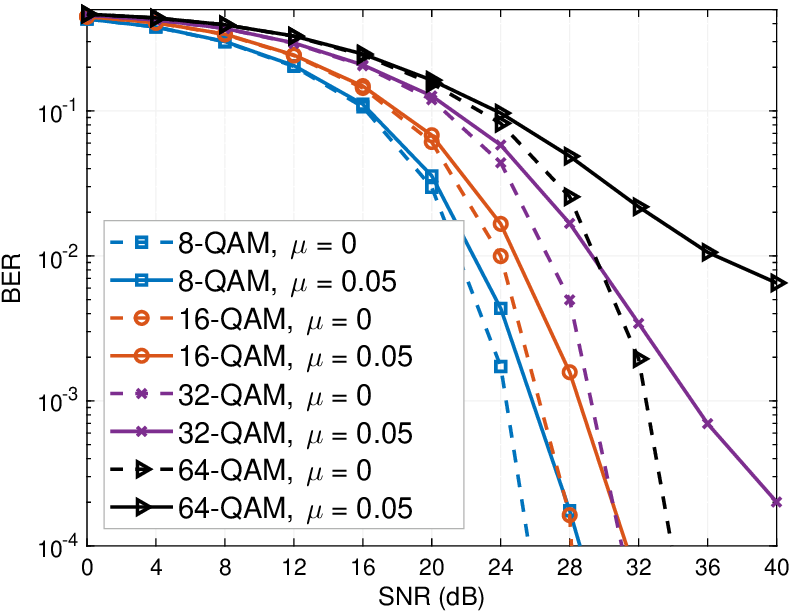}
   \caption{After CPE correction}
    \label{QAM2}
\end{subfigure}\hfill
 \caption{BER versus SNR for DCO-OFDM with different QAM constellation orders at
  normalized Doppler $\mu \in \{0,\,0.05\}$.}
\label{fig:QAM}
\end{figure*}

Given the model in \eqref{eq:freqmodel}, one approach to compensate the CPE is to estimate it and to re-align the received signal constellation accordingly before baseband demodulation. Here, we propose to perform per-symbol CPE estimation using a set of pilot subcarriers $\mathcal{P}$ using which the estimated phase is:
\begin{equation}
    \hat{\phi}=\arg \left(\sum_{p \in \mathcal{P}} Y_{s}[p] X_s^*[p]\right),
\end{equation}
where $.^*$ denotes complex conjugate. Then, the CPE can be corrected as $\breve{Y}[k]={Y}_s[k]\, e^{-j \hat{\phi}}$, where the compensated symbols $\breve{Y}[k]$ can be passed to a standard symbol demapper.

\begin{figure}[!t]
  \centering
  \includegraphics[width=0.95\linewidth, height=6cm]{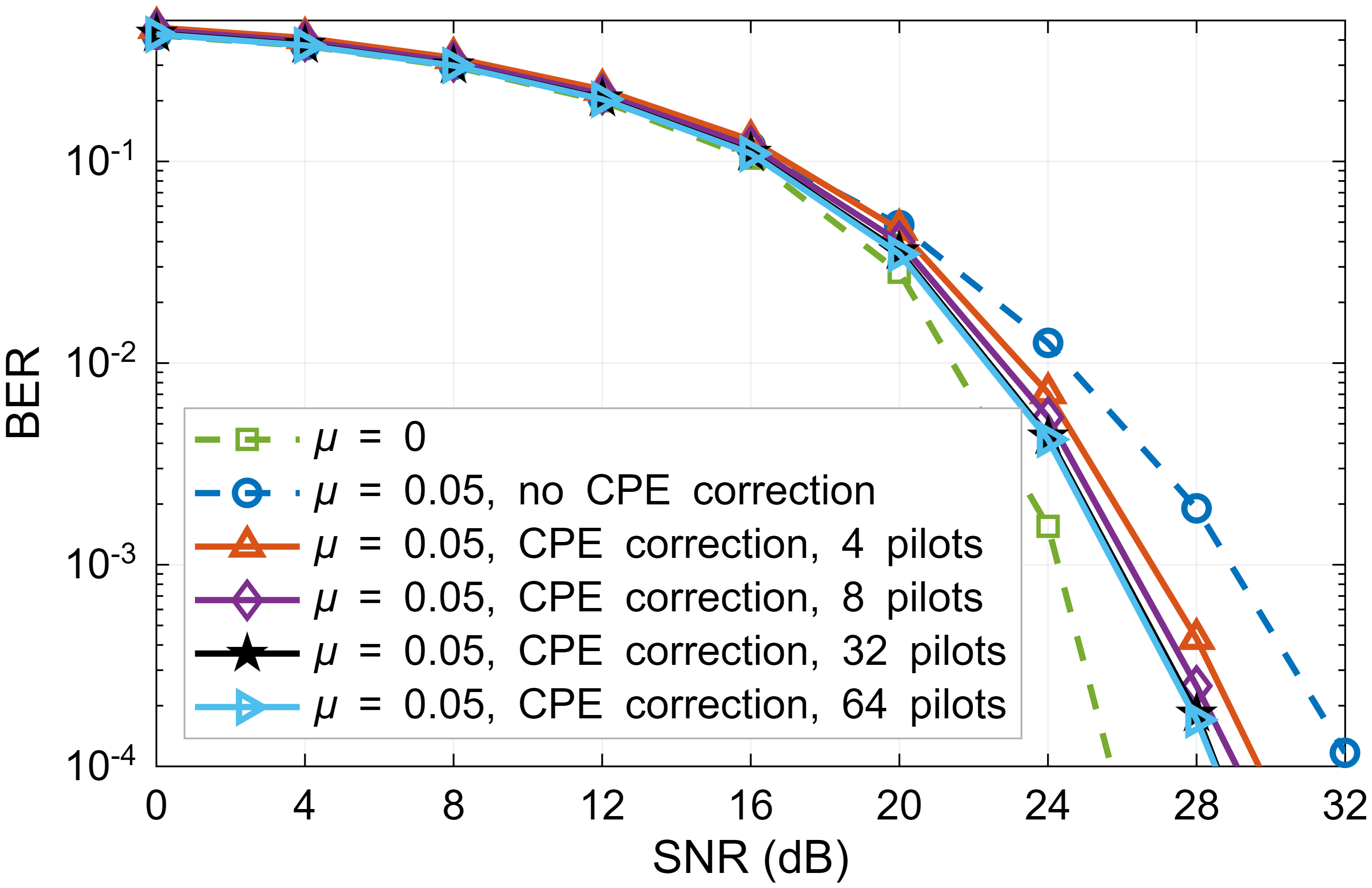}
 
  \caption{BER versus SNR for the case of 8-QAM DCO-OFDM, under no Doppler ($\mu=0$) and Doppler shift (with $\mu=0.05$) with CPE correction using different numbers of pilot subcarriers $|\mathcal{P}|$.}
  \label{fig:pilot}
\end{figure}

\section{Numerical Results}
\label{sec:results} 
We present numerical simulations to assess the impact of Doppler on the bit-error-rate (BER) performance of FSO links as well as the efficiency of the proposed CPE tracking method.

\subsection{Simulation Parameters} \label{sec:SimulationParameters}
We consider DCO-OFDM signaling with $N{=}4096$ subcarriers and Gray-mapped power-normalized QAM symbols. We reasonably neglect channel delay spread and therefore omit the cyclic prefix, isolating hence the impact of Doppler and ICI. 
A set of $32$ pilot subcarriers per OFDM symbol is used for CPE estimation and correction. We consider normalized Doppler values of $\mu = f_D / \Delta f \in \{0.05, 0.1, 0.2\}$. 
The first two values of $\mu$ correspond to representative LEO–OGS links and LEO ISL scenarios, respectively (see Appendix for the calculation details), while the larger value $\mu = 0.2$ is included to illustrate more severe Doppler conditions and to better highlight the observed trends in the BER performance.
\subsection{Illustration of Doppler Effect on Signal Constellation} \label{res:constellation}
Figure~\ref{R2} shows how Doppler affects the baseband constellation of a 4-QAM DCO-OFDM system, with the transmitted constellation shown in Fig.\,2(a). As seen in Fig.\,~\ref{O2b}, when no Doppler shift is present, only the effect of Rx noise is observed.
For $\mu=0.1$ (Fig.\,~\ref{O2c}), the Doppler shift manifests as a phase rotation of QAM clusters, accompanied by increased spreading due to ICI. Increasing the Doppler to $f_D/\Delta f = 0.2$ (Fig.\,\ref{O2d}) results in further phase rotation and a stronger ICI, which causes more pronounced cluster smearing. Lastly, 
Figs.~\ref{O2e} and~\ref{O2f} show the constellations after applying the CPE correction proposed in Section~\ref{sec:rx-model}. We notice that this simple correction largely eliminates the CPE, while the constellation spreading due to ICI remains almost unchanged.

\subsection{BER Performance} \label{res:ber}
To contrast the CPE and ICI effects, Figs.~\ref{BER1} and \ref{BER2} show the BER performance under Doppler shift without and with per-symbol CPE correction, respectively.
As seen, without CPE correction (Fig.~\ref{BER1}),
increasing $\mu$ progressively degrades the BER. 
The CPE correction in Fig.~\ref{BER2} substantially recovers the performance loss at relatively small Doppler values, i.e., at $\mu=0.05 ,\ 0.1$; the remaining BER degradation is due to residual ICI. At higher Doppler shifts, i.e., $\mu = 0.2$, the residual ICI is significant and results in a BER floor. Note that eliminating the BER floor requires ICI-aware equalization based on a time-varying channel model~\cite{barhumi-TSP-2006}.

The CPE and ICI effects become more pronounced at higher constellation orders, as shown in Fig.~\ref{QAM1} for the normalized Doppler $\mu = 0.05$ (representative of LEO--to--OGS links), together with the corresponding no-Doppler BER benchmarks. After CPE correction (Fig.~\ref{QAM2}), a substantial BER improvement is obtained; however, for higher-order QAM, a BER floor remains due to residual ICI. Figure~\ref{fig:pilot} further illustrates the impact of the number of pilot subcarriers on the proposed CPE correction method for the case of 8-QAM. As observed, while increasing the pilot count enhances the CPE estimation accuracy and improves the BER performance, the gain becomes marginal beyond a moderate number of pilots.

\section{Discussion and Conclusion}
\label{sec:disc}
In multiple-subcarrier IM/DD OWC systems employing, for instance O-OFDM or O-OTFS signaling, the transmit waveform is made
real by imposing Hermitian symmetry on the subcarrier mapping. 
At the Rx, the baseband-equivalent signal phase, i.e., the phase of the received subcarrier
coefficients $Y[k]$ after the DFT, contains both the complex symbols' phase and the frequency-dependent channel effects (e.g., Doppler and misalignment-induced frequency offset). 
In other words, although direct signal detection makes the link insensitive to the optical-carrier phase, it does not eliminate the baseband-equivalent (electrical-domain) phase of the subcarriers. Channel effects modeled as time shift, time-scaling, or frequency offset of the real time-domain waveform, appear in the DFT domain as complex exponentials, inducing constellation rotation and subcarrier mixing, i.e., CPE and ICI. These baseband phase terms must
therefore be estimated and equalized, e.g., using pilot-aided CPE tracking, especially in high-Doppler scenarios such as in ISLs and
LEO–to–OGS links. Note, this is in contrast with single-subcarrier IM/DD links which are not sensitive to baseband-equivalent phase.

\appendix

\section*{Doppler Shift in LEO-OGS Links and ISLs} 
For a LEO to ground link, the Doppler shift is given by~\cite{Chen-PTL-2025}: 
\begin{equation} \label{eq:D-shift}
f_{D} = -\frac{f}{c}\,
\frac{\omega_s R_E R_S \sin\phi\,\eta(\phi_{\max})}
{\sqrt{R_E^2 + R_S^2 - 2 R_E R_S \cos\phi\,\eta(\phi_{\max})}},
\end{equation}
where $c$ is the speed of light, $R_E$ and $R_S$ denote the Earth radius and satellite orbital radius, respectively, $\phi_{\max}$ is the Earth-center edge-of-visibility angle given by $\phi_{\max}=\cos^{-1}(R_E/R_S)$, and $\omega_s$ denotes the satellite angular rate, given by $\omega_s=\sqrt{GM/R_S^3}$, with $GM$ being the gravitational parameter. Also,
\begin{equation} \label{eq:eta}
\eta(\phi_{\max}) =
\cos\!\Big[\cos^{-1}\!\Big(\frac{R_E}{R_S}\cos\phi_{\max}\Big)-\phi_{\max}\Big].
\end{equation}
In~\eqref{eq:D-shift}, $f$ denotes the frequency at which Doppler is evaluated, i.e., $f=f_m$ for the IM/DD electrical domain. As noted in Section~\ref{Subsec-Doppler-Mod}, we evaluate the Doppler shift $f_D$ at the nominal band-edge frequency $f_{m,\mathrm{e}}=(N/2-1)\Delta f$ (worst case). Hence, $f=f_{m,\mathrm{e}}=(N/2-1)\Delta f$. \\
We evaluate $f_D$ at $\phi=\phi_{\max}$ and further define
\begin{equation*}
\alpha^\prime \triangleq
\frac{\omega_s R_E R_S \sin\phi_{\max}\,\eta(\phi_{\max})}
{c\sqrt{R_E^2 + R_S^2 - 2 R_E R_S \cos\phi_{\max}\,\eta(\phi_{\max})}}.
\end{equation*}
Then, \eqref{eq:D-shift} can be written as
\begin{equation*}
f_D = -\alpha^\prime (N/2-1)\Delta f.
\end{equation*}
Considering $R_E=6378$~km, $GM=3.986004418\times10^{14}$~m$^3$/s$^2$, and a typical LEO altitude $h=500$~km, we obtain $R_S=R_E+h=6.878\times10^6$~m, $\phi_{\max}=0.384$~rad, $\eta(\phi_{\max})\approx0.9885$, and $\alpha^\prime\approx2.18\times10^{-5}$. \\
Setting $\Delta f=50$~kHz, we finally obtain $f_D=-2.23\times10^3$~Hz (the negative sign indicates an approaching satellite). Hence, the resulting normalized Doppler is $\mu = f_D/\Delta f= 0.045$.

For ISLs,  Doppler shift can be approximated from the relative velocity $v_{\text{rel}}$ between the two satellites. {For the $500$~km orbit considered here, the orbital speed is $
{v_{\text{orb}}} = \sqrt{{GM}/{R_S}}\approx 7.61~\text{km/s}$. Two counter-rotating satellites on this orbit have the relative velocity of $v_{\text{rel}} \approx 2\, v_{\text{orb}}
\approx 15.2~\text{km/s}$. Using \mbox{$f_{D} \approx f_{m,\mathrm{e}} v_{\text{rel}}/{c} $}, 
the corresponding normalized Doppler equals \mbox{$\mu={f_{D
}}/{\Delta f} \approx (N/2-1)\,{v_{\text{rel}}}/{c} \approx 0.1$}.}%

\section*{Acknowledgment}
This work  has received funding from the European Union’s Horizon Europe research and innovation programme under the Marie Skłodowska-Curie Doctoral Network “FOCAL” (Grant Agreement No. 101169042).

\balance

\bibliographystyle{IEEEtran}
 \bibliography{IEEEabrv,Sources}

\balance

\end{document}